\documentclass[aps,preprint,preprintnumbers,amsmath,amssymb,superscriptaddress,longbibliography]{revtex4-1}

\usepackage{graphicx}
\usepackage{dcolumn}
\usepackage{bm}

\newcommand{\sss}{\mathrm{s}}
\newcommand{\bb}{\mathrm{B}}
\newcommand{\bbb}{\mathrm{b}}

\begin{document}

\title{Pushing the  glass transition towards random close packing using  self-propelled hard spheres}

\author{Ran Ni}
\email{rannimail@gmail.com}
\affiliation{%
Laboratory of Physical Chemistry and Colloid Science, Wageningen University, Dreijenplein 6, 6703 HB Wageningen, The Netherlands
}%
\affiliation{%
Van $'$t Hoff Institute for Molecular Sciences, Universiteit van Amsterdam, Science Park 904, 1098 XH Amsterdam, The Netherlands
}%

\author{Martien A. Cohen Stuart}
\affiliation{%
Laboratory of Physical Chemistry and Colloid Science, Wageningen University, Dreijenplein 6, 6703 HB Wageningen, The Netherlands
}%

\author{Marjolein Dijkstra}
\affiliation{Soft Condensed Matter, Utrecht University, Princetonplein 5, 3584 CC Utrecht, The Netherlands}

\begin{abstract}
Although the concept of random close packing with an almost universal packing fraction
of $\sim 0.64$ for hard spheres was introduced more than half a century ago, there are
still ongoing debates. The main difficulty in searching the densest packing is that states with
packing fractions beyond the glass transition at $\sim 0.58$ are inherently non-equilibrium systems, where the dynamics slows down with a structural relaxation time diverging with density; hence,
the random close packing is inaccessible. Here we perform simulations of self-propelled hard spheres, and we find that
with increasing activity the relaxation dynamics can be sped up by orders
of magnitude. The glass transition shifts to higher packing fractions upon increasing the activity,
allowing the study of sphere packings with fluid-like dynamics at packing fractions close to random close packing. Our study opens new possibilities of investigating dense packings and the glass transition in systems of hard particles.
\end{abstract}

\maketitle
Random packings of hard spheres have been extensively studied as models for liquids, glasses, colloids, granular systems, and living cells, but they are also related to many mathematical problems, such as signal digitalization, error correcting codes, and optimization problems~\cite{rmp2010}.
Nevertheless, there are  ongoing debates on the characterization of amorphous packings of  hard spheres, and the nature of random close packing (RCP) is still highly controversial~\cite{bernal1960,torquato2000,ohern2002,kamien2007}. One of the main issues is whether or not RCP is well-defined with a single value for the jamming density, also called the J-point, or by a finite range of jamming densities, a so-called J-line~\cite{rmp2010}. For instance, it has been argued that RCP is ill-defined as one can always obtain a higher packing fraction by introducing crystalline order in the system \cite{torquato2000}. Moreover,  experiments and simulations show that a  range of jamming densities  can be obtained depending on the protocol that is used to generate random sphere packings \cite{ohern2002}. The observed range of jamming densities has been explained by the rate at which accessible states disappear \cite{kamien2007}, and by the theoretical prediction that the jammed configurations can be regarded as the infinite-pressure limit of glassy states \cite{biazzo2009,mari2009}. Subsequent simulations indeed confirmed  that a system falls out of equilibrium at a density where the relaxation time exceeds the compression rate, and that this glass phase can only be compressed further to its corresponding jammed state as structural relaxation is prohibited \cite{berthier2009,hermes_epl_2010,chaudbury2010}. Hence, a  range of different jamming densities can thus be obtained as the glass transition depends on compression rate.

The dynamics of self-propelled (active) particles in dense suspensions has attracted an increasing amount of interest in order to describe self-organization phenomena like bird flocks, bacteria colonies, tissue repair, and cell cytoskeleton. Very recently, also  artificial colloidal microswimmers have become available, which show promise for applications such as biosensing, drug delivery, etc.~\cite{ebbens2010}.
Examples of active self-propelled particles are colloids with magnetic beads that act as artificial flagella \cite{dreyfus2005}, catalytic Janus particles \cite{howse2007,erbe2008,palacci2010,baraban2012},  laser-heated metal-capped particles \cite{volpe2011}, light-activated catalytic colloidal surfers \cite{palacci2013}, platinum-loaded stomatocytes \cite{wilson2012}, or living bacteria \cite{schwarz-linek2012}. In contrast to passive colloidal particles that only exhibit Brownian motion due to thermal fluctuations, active self-propelled colloids are driven by the interplay between random fluctuations and active swimming. While passive Brownian particles are in thermodynamic equilibrium with the solvent, active particles are driven far from  equilibrium as they  convert incessantly energy into active motion. Recent experiments on crowded systems of active colloids and active cells show dynamical features like jamming and dynamical arrest that are very  similar to those observed in  glassy materials \cite{weitz2010,palacci2010}. These experimental observations are in line with recent theoretical predictions \cite{berthier2013} that dynamic arrest can occur in systems that are far from equilibrium such as active and driven systems with characteristics that are very similar to those observed in glasses. Additionally, it was predicted theoretically that the non-equilibrium glass transition moves to lower temperature with increasing activity and to higher temperature with increased dissipation in spin glasses~\cite{berthier2013}.
However, the temperature shift of the non-equilibrium glass transition could not be explained by  replacing the temperature with an effective temperature~\cite{berthier2013}.

Inspired by these results, here we investigate the dynamical behavior of dense suspensions of active particles and study whether or not a glass transition can occur in a system that is far out of equilibrium.  To exclude the effect of temperature, we investigate the effect of self-propulsion on dense packings of hard spheres that solely interact with excluded volume interactions. Remarkably, the glass transition shifts to higher packing fractions upon increasing the activity, which allows us to study for the first time random sphere packings with fluid-like dynamics at volume fractions close to RCP. We also find that the effect of random self-propulsions decreases with larger rotational diffusion coefficients of the particles. Moreover, we confirm the existence of the non-equilibrium glass transition in active systems with a realistic particle model for the first time. 

\section*{Results}
\subsection*{Non-equilibrium glass transition}
\begin{figure}[ht!]
\includegraphics[width = 0.8\textwidth]{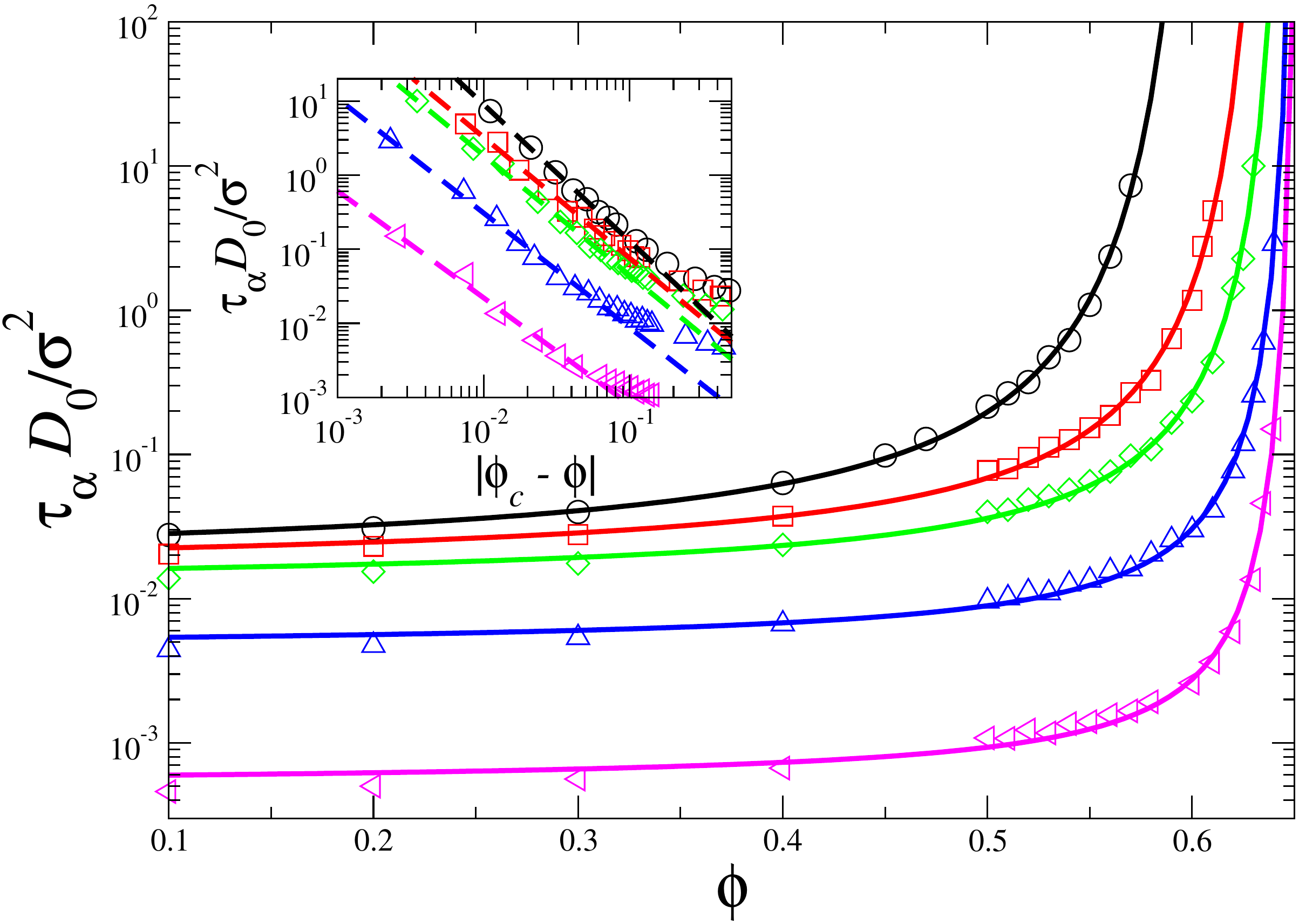}
\caption{\label{fig1} {\bf Relaxation time of self-propelled hard-sphere systems.} Structural relaxation time $\tau_\alpha$ as a function of packing fraction $\phi$ for hard-sphere systems with various self-propulsions $f$. Symbols: circles, squares, diamonds, triangles, and left-pointing triangles are results for $f\sigma / k_\bb   T = 0,10,20, 80$ and 800, respectively. The solid lines are the VFT fits: $\tau_\alpha \sim \exp[B\phi_0 / (\phi_0 - \phi)]$, and in the inset the dashed lines denote the MCT fits: $\tau_\alpha \sim |\phi_\mathrm{c} - \phi|^{-\gamma}$. The VFT and MCT fitting parameters are tabulated in Table~\ref{tab1}.}
\end{figure}

We perform event-driven Brownian Dynamics (EDBD) simulations ~\cite{scala2007,scala2013} of  $N=2000$ self-propelled hard spheres with a Gaussian size polydispersity of
$8\%$ and a mean diameter of $\sigma$.  The self-propulsion force for particle $i$ is modelled as a constant force $f$ with an orientation $\mathbf{\hat{u}}_i(t)$, which undergoes free Brownian rotation~\cite{lowen2012}. We employ $D_0\sigma^{-1}$ as our unit of velocity in order to have a direct comparison of the self-propulsion force to the thermal Brownian motion. To relate to experiments, one can employ the Stokes-Einstein relation with $D_0=k_\bb   T/3 \pi \eta \sigma$  and  $\eta$ the viscosity of the solvent. Note that the self-propulsion speed of a single particle is given by $f \sigma/k_\bb T \times D_0\sigma^{-1}$. To prepare the initial  configuration, we use the Lubachevsky-Stillinger algorithm~\cite{LSA} to grow the particles in the simulation box to the packing fraction of interest. Subsequently, we ``equilibrate'' the system until each particle has moved on average more than several times  its own diameter, and the system has reached a steady state. We confirm that there is no crystalline order observed in all simulations. To investigate the dynamical relaxation of the system, we measure the self-intermediate scattering function  $F_\sss(q,t) = 1/N \sum_{j=1}^{N} {\left \langle \exp \left \{i \mathbf{q} \cdot \left[\mathbf{r}_{j}(0) - \mathbf{r}_{j}(t) \right] \right\} \right \rangle}$,
where $\mathbf{q}$ is the wave vector for which we use $q = |\mathbf{q}| = 2 \pi /\sigma$.  Figure~\ref{fig1} shows the structural relaxation time $\tau_{\alpha}$, defined by $F_\sss(q,\tau_\alpha) = e^{-1}$,  as a function of packing fraction $\phi$ for self-propulsion $f \sigma/k_\bb   T=0$ (passive spheres), 10, 20, 80, and 800. We clearly observe that  the dynamics slows down with orders of magnitude upon increasing the packing fraction for both  passive and active systems. In addition, we find that the structural relaxation time of both passive and active systems can be fitted  with the Vogel-Fulcher-Tammann (VFT) law, $\tau_\alpha \sim \exp\lbrack B \phi_0/(\phi_0-\phi) \rbrack$ and the mode coupling theory (MCT) prediction, $\tau_\alpha \sim |\phi_\mathrm{c}-\phi|^{-\gamma}$ \cite{brambilla2009}. For a passive hard-sphere system, we find in agreement with Refs. \cite{brambilla2009,rmp2010} that the structural relaxation  $\tau_{\alpha}$  can be described by an algebraic MCT divergence at a packing fraction  $\phi_\mathrm{c} \approx 0.58$, and an exponential VFT divergence  at $\phi_0 \approx 0.62$. For active hard spheres,  we observe  that $\tau_{\alpha}$ decreases by orders of magnitude upon increasing the self-propulsion $f$ at fixed $\phi$. However, for sufficiently high densities the dynamics slows down dramatically and the structural relaxation time is again well-described by the MCT and VFT predictions as shown in Fig.~\ref{fig1}.   We thus find that the dynamical relaxation behavior of active hard spheres is very similar to that of passive hard spheres, as our results show clear evidence for dynamical arrest and a diverging structural relaxation time even for systems that are far out-of-equilibrium at very high activity.

\begin{figure}[ht!]
\includegraphics[width = 0.8\textwidth]{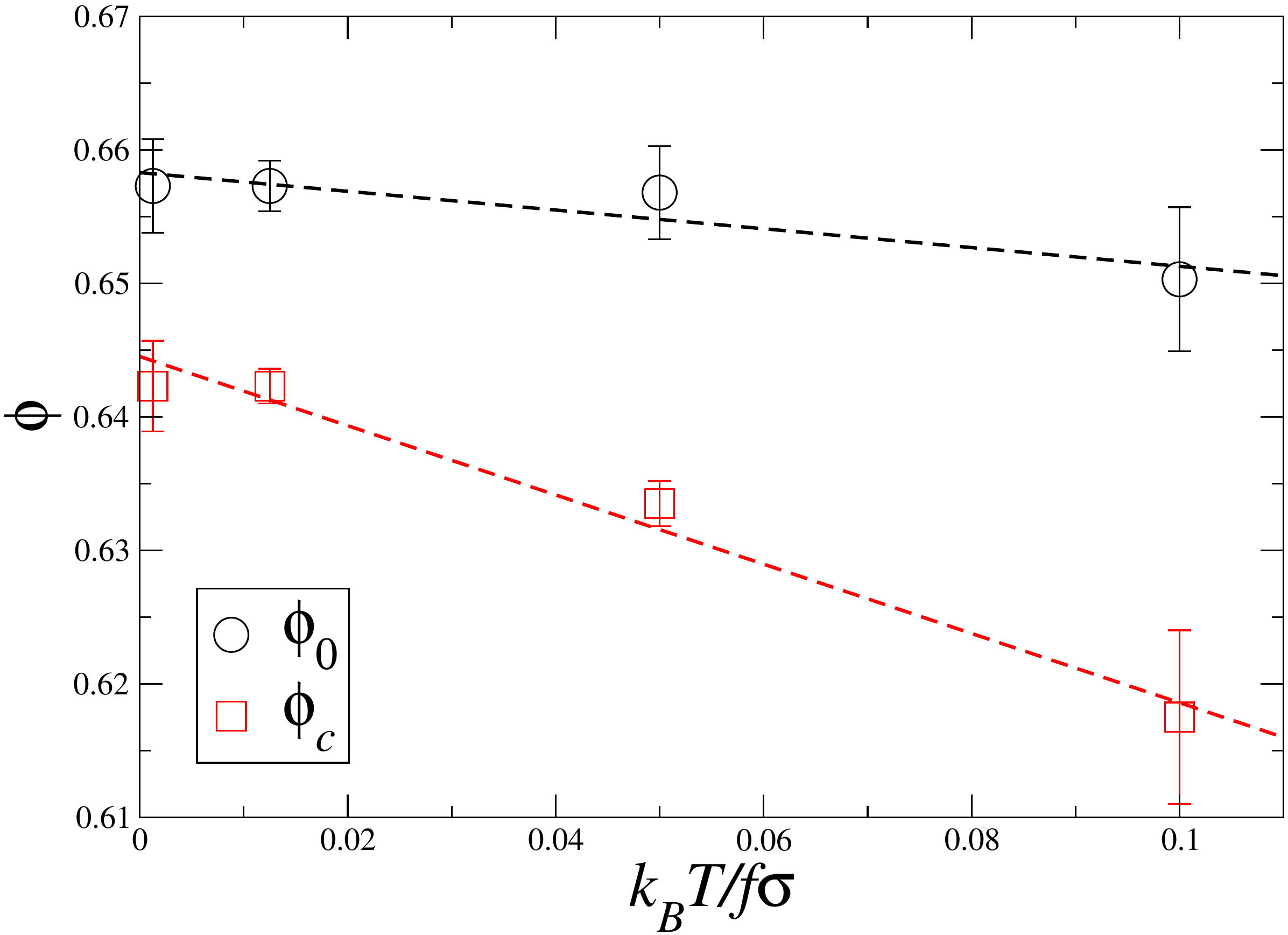}
\caption{\label{fig2} {\bf Glass transition packing fractions of self-propelled hard-sphere systems.} The critical packing fractions $\phi_0$ and $\phi_\mathrm{c}$ as obtained from  the VFT and MCT fits, respectively, of the structural relaxation time $\tau_\alpha$ for active hard spheres as a function of the inverse  self-propulsion $k_\bb  T/f \sigma$. The dashed lines are the best linear fits.}
\end{figure}

In Table~\ref{tab1}, we present the resulting critical packing fractions $\phi_0$ and $\phi_\mathrm{c}$ as obtained from the VFT and MCT fits, respectively, as well as the critical exponent $\gamma$ and prefactor  $B$. We find that $\phi_\mathrm{c}$ as obtained from the MCT fits increases monotonically with self-propulsion $f$.  For passive hard spheres, we find an MCT  critical exponent  $\gamma=1.844 \pm 0.551$, which agrees with previous reported values of $\gamma \in [1.7,2.5]$ \cite{kumar2006,voigtmann2004}. In addition, we note that the critical exponent $\gamma$   decreases slightly with increasing $f$, but we also mention here that $\gamma$ is extremely sensitive to the precise $\phi$-range that is used in the MCT fitting as also noted in Refs. \cite{kumar2006,voigtmann2004}. Similarly, by fitting the structural relaxation time $\tau_\alpha$ with the VFT prediction, we find that  $\phi_0>\phi_\mathrm{c}$ also increases monotonically with self-propulsion $f$, but less pronounced than $\phi_\mathrm{c}$. We also note that $\phi_0$ reaches a constant value $\phi_0 \simeq 0.6573$ within our statistical accuracy for $f \sigma/k_\bb   T >20$.
Figure \ref{fig2} shows  $\phi_0$ and $\phi_\mathrm{c}$ as a function of the inverse activity $1/f$. We find that the packing fraction range $\lbrack \phi_\mathrm{c},\phi_0\rbrack$ for which the dynamical relaxation is characterized by an exponential VFT divergence decreases  with decreasing $1/f$, but remains finite even in the limit of infinite activity, i.e., $1/f \rightarrow 0$. This also suggests that even with infinitely strong random self-propulsions, during the process of producing RCP, the system yet undergoes an MCT glass transition, in which the structural relaxation time diverges with a power law, and the obtained RCP of hard spheres still depends on the speed of compression. 

We also observe that the prefactor $B$ as determined from the VFT fits decreases with activity $f$, which corresponds to a stronger dependence of $\tau_\alpha$ with packing fraction $\phi$. The $\phi$-dependence of $\tau_\alpha$  allows us to classify the various glasses in terms of their ``fragility''. The concept of ``fragility'' was originally introduced for molecular glasses to describe the sensitivity of $\tau_\alpha$ to changes in temperature: Fragile glasses show a strong temperature-dependence of $\tau_\alpha$, whereas  strong glass-formers correspond to a weak sensitivity to  temperature. The analogy between volume-dependent fragility in colloidal glasses and the temperature-dependent fragility in molecular glasses, where $T$ is exchanged with $1/\phi$, has  been demonstrated recently in experiments on deformable colloidal particles, where softer and more deformable particles lead to stronger glass behavior compared to hard particles \cite{mattsson2009}. We thus find from our VFT fits that $\tau_\alpha$ depends   stronger  on packing fraction $\phi$ upon increasing $f$ as the prefactor $B$ decreases, which corresponds to more fragile glass behavior.

The fragility is related to the structural relaxation in glasses, which can be described by activated processes like cage-breaking or barrier hopping in a complex energy landscape that reflects the possible glass configurations. Upon increasing packing fraction $\phi$, the barriers between local minima in the energy landscape that correspond to different glass configurations increase and the system gets kinetically trapped in a local minimum at short times, and  jumps only to other minima on a longer time scale due to activated dynamics. Assuming that the relaxation time is determined by the barrier height of the activated process, one can determine the effective  activation energy (or entropic barrier height) $\Delta F_\bbb$   from the VFT fits using $\Delta F_\bbb \simeq B \phi_0/(\phi_0-\phi)$. In the case of passive hard spheres, we find that  $\Delta F_\bbb \simeq 8 ~k_\bb   T$ at $\phi =0.58$, which agrees well with the theoretical predictions of Ref. \cite{schweizer2003}. However, the entropic barrier $\Delta F_\bbb$ for passive hard spheres as obtained from the VFT fits increases  rapidly with $\phi$, and reaches a value of $\Delta F_\bbb \simeq 30 k_\bb   T$ at $\phi =  0.61$, which is much higher than the theoretically predicted barrier height of $\sim 14 k_\bb  T$ in  Ref. \cite{schweizer2003}, and $\sim 16 k_\bb  T$ as obtained from the experimental observations and analysis of the structural rearrangements in colloidal glasses under shear \cite{schall2007}. A possible explanation for the sharp increase in relaxation time $\tau_\alpha$ and barrier height $\Delta F_\bbb$ with increasing packing fraction $\phi$ is that a larger cooperativity of the particles is required in order to cross the barrier. For self-propelled spheres, we find that the effective activation energy $\Delta F_\bbb$ at $\phi=0.61$ decreases to only a few $k_\bb  T$ for self-propulsions $f \sigma/k_\bb T>20$. Hence, the particles can easily escape out of their cages  due to the activated dynamics of the self-propulsions, which is consistent with the fact that $\phi=0.61$ is below the MCT critical packing fraction $\phi_\mathrm{c}$.

\begin{figure*}
\includegraphics[width = 0.9\textwidth]{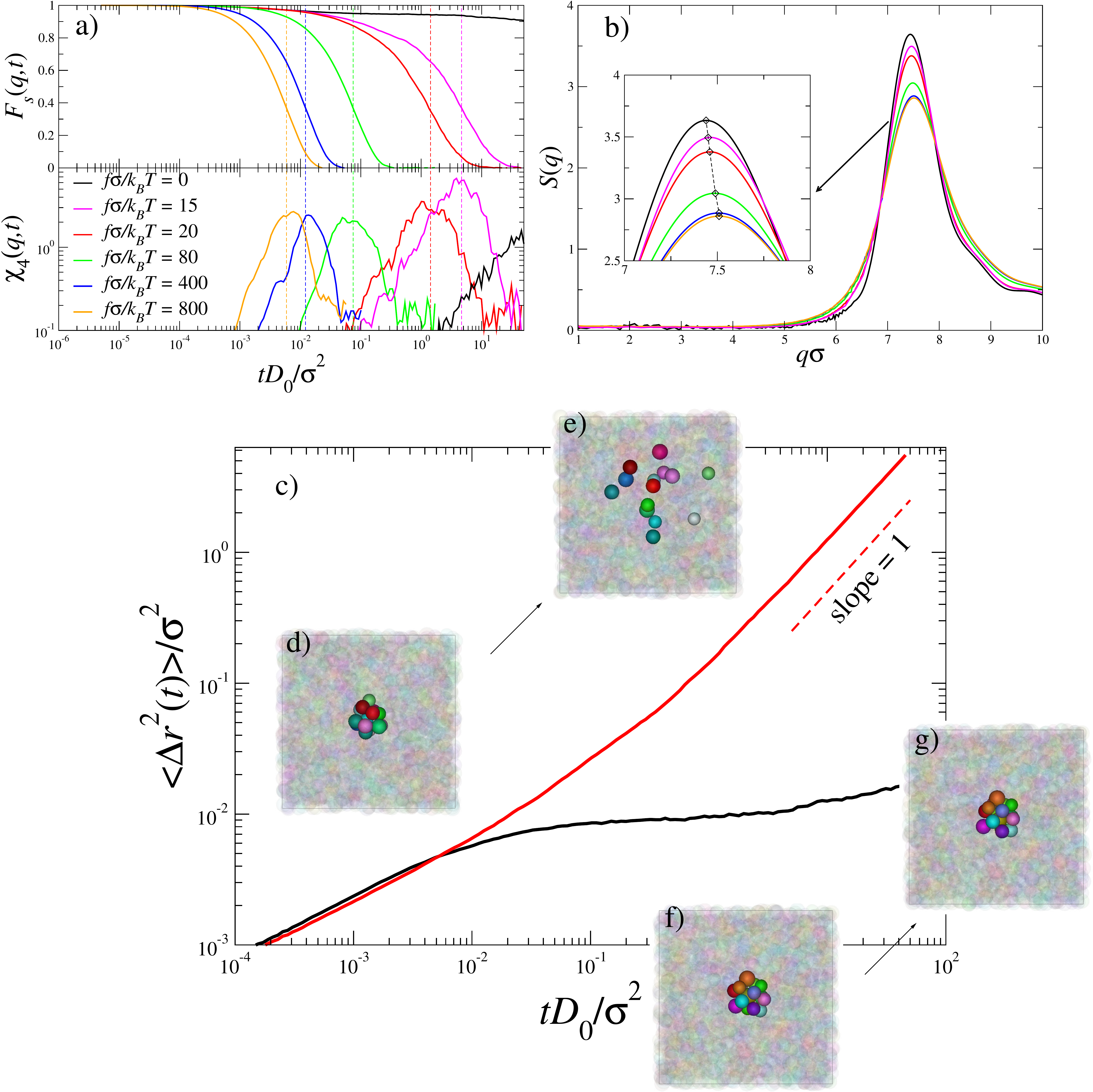}
\caption{\label{fig3} {\bf Effect of self-propulsions on the structural relaxation.} (a) The self-intermediate scattering function $F_\sss(q,t)$, and  the four point dynamic susceptibility $\chi_4(q,t)$, computed at wave vector $q = 2\pi/\sigma$ as a function of time $tD_0/\sigma^2$ in systems of active  hard spheres at packing fraction  $\phi = 0.62$ and varying self-propulsions $f \sigma/k_\bb   T$ as labeled. The dashed vertical lines denote the location of structural relaxation times $\tau_{\alpha}$. (b) The structure factor $S(q)$ of the systems as described in (a) with an enlarged view of the maxima of the peaks as  denoted by the open diamonds in the  inset. (c) The mean square displacement $\langle r^2(t) \rangle$ as function of time $tD_0/\sigma^2$ for a system of active hard spheres at $\phi=0.62$ with $f\sigma/k_\bb  T = 0$ (black line) and 20 (red line), respectively. (d,e) and (f,g) are snapshots from a typical dynamic trajectory of a system with $f\sigma/k_\bb  T = 0$ and 20, respectively, in which only a randomly selected particle and its contacting neighbours are shown as solid spheres while other particles are semi-transparent. (d,f) and (e,g) are at a time $tD_0/\sigma^2 = 0$ and 50, respectively.}
\end{figure*}

For active hard spheres, we observe that the relaxation time decreases by orders of magnitude as it becomes much easier for self-propelling motorized hard spheres to cross the barriers and to explore different glass configurations than for passive spheres.  To better understand the effect of self-propulsion on the dynamics of hard-sphere glasses, we plot  the self-intermediate scattering function $F_\sss(q,t)$ for a system of active hard spheres   with varying  self-propulsions $f$ at  $\phi = 0.62$ in Fig.~\ref{fig3}a.  We observe that $F_\sss(q,t)$ reaches a cage-trapping plateau and remains stuck in an amorphous configuration in the case of passive hard spheres, whereas for active spheres  $F_\sss(q,t)$ decays to zero within our simulations, and  the relaxation time  decreases significantly with self-propulsion. The common picture of glassy dynamics
and barrier crossings is that a particle rattles for a long time inside a cage that is formed
by its neighbours and then suddenly breaks out of its cage.  To investigate whether or not the barrier crossing corresponds to an abrupt cage-breaking event and how localized these events are, we measure the four-point dynamic susceptibility, $\chi_4(q,t) = N \langle \delta F_\sss^2(q,t) \rangle$, where $\delta F_\sss(q,t)$ denotes the fluctuating part of $F_\sss(q,t)$, as shown in Fig.~\ref{fig3}a.  As expected,  $\chi_4(q,t)$ increases initially in time and shows a peak on a time scale that coincides with the structural relaxation time $\tau_{\alpha}$ before it decays  at longer times. The height of the peak of $\chi_4(q,t)$ is related to the average number of particles that are dynamically correlated in the structural relaxation process \cite{brambilla2009,berthier2007}.  Figure~\ref{fig3}a shows that the peak height of $\chi_4(q,t)$  is $\simeq 2$ for $f\sigma/k_\bb  T \ge 80$, providing support that there is almost no collective motion involved in the structural relaxation, which is to be expected as $\phi=0.62$ is far below the MCT critical packing fraction $\phi_\mathrm{c}$ at $f\sigma/k_\bb  T \ge 80$.  In addition, we find clearly that the peak height increases upon decreasing the self-propulsion towards $f=0$ and thus also decreasing MCT critical packing fraction $\phi_\mathrm{c}$. We note however that we were not able to determine the peak height at $f=0$ as the relaxation time exceeds our simulation time.  We thus find that even for self-propelled hard spheres the structural relaxation becomes more cooperative  for packing fractions close to the MCT glass transition, i.e., $\phi \rightarrow \phi_\mathrm{c}$.

\begin{figure}[ht!]
\includegraphics[width = \textwidth]{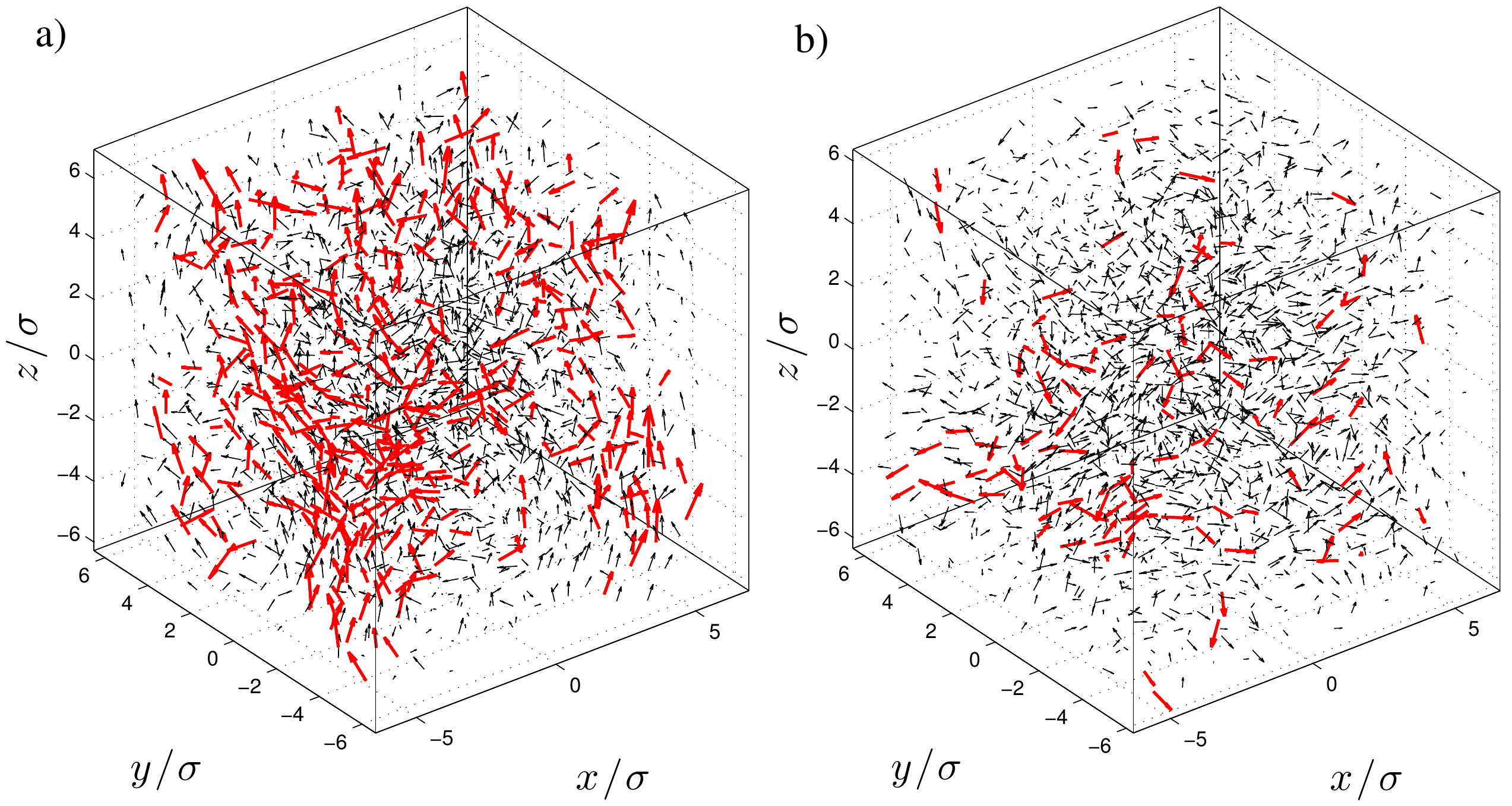}
\caption{\label{fig4} {\bf Structural relaxation of self-propelled hard-sphere systems.} The distribution of displacements in a system of active hard spheres at $\phi = 0.62$ with  self-propulsion $f\sigma/k_\bb  T = 20$ (a) and 80 (b) measured over a time interval $\simeq \tau_{\alpha}$. The red thick arrows highlight the displacements larger than $0.7\sigma$.}
\end{figure}

To corroborate our finding on the cooperative nature of the structural relaxation for self-propelled spheres, we show in  Fig.~\ref{fig4} the distribution of displacements
in a system of active hard spheres at $\phi = 0.62$ measured over a time window that spans the structural relaxation time $\tau_{\alpha}$. Figure~\ref{fig4} clearly shows that for  $f\sigma/k_\bb  T = 20$  the fast moving particles, i.e., particles with displacements larger than $0.7\sigma$, are spatially correlated, which is to be expected as the system is close to the MCT glass transition. We wish to  mention here that collective motion was also observed in   recent simulations  on a 2-dimensional active hard-disk system~\cite{berthier2013prl}. For larger self-propulsions, i.e. $f\sigma/k_\bb  T = 80$,  the motion of the particles is less correlated as the system is well below $\phi_\mathrm{c}$. Hence,  the dynamics becomes more fluid-like and the dynamic heterogeneities disappear in a system of active hard spheres with $f\sigma/k_\bb  T = 80$ and $\phi=0.62$.

We also study the effect of  self-propulsion on the cage dynamics.  To this end, we measure the mean square displacement $\langle \Delta r^2(t) \rangle$ in a system of passive hard spheres at  $\phi=0.62$. The mean square displacement as shown in Fig. \ref{fig3}c shows a clear plateau at long time scales as the particles  are kinetically trapped within the cages that are formed by their neighbours. To illustrate the cage effect, we present snapshots  of a typical dynamical trajectory in Fig.~\ref{fig3}f-g, where we highlight the  particles that form the cage. We clearly observe that the cage remains intact on at least a time scale of $tD_0/\sigma^2 = 50$ with $D_0$ the mean short-time diffusion coefficient. For comparison, we also plot  $\langle \Delta r^2(t) \rangle$ for a system of active hard spheres with $f \sigma/k_\bb  T=20$, which shows diffusive behavior at short times, and subdiffusive behavior at intermediate time scales but without a clear cage-trapping plateau. At long time scales, $\langle \Delta r^2(t) \rangle$  grows  linearly with time and the particles become diffusive again with a long time diffusion coefficient $D_{\mathrm{L}}/D_0 \simeq 0.02$. The snapshots in Fig.~\ref{fig3}d-e show that the cage has disappeared completely within a time scale of  $tD_0/\sigma^2 = 50$, which explains the absence of any cage effect in $\langle \Delta r^2(t) \rangle$ (see Supplementary Movie 1 and 2).

Finally, we investigated  if a particle can break out of its cage by its own self-propulsion. We performed a simulation of a single active particle with activity $f \sigma/k_\bb   T=80$ in a system of $N=2000$ hard spheres at $\phi=0.62$, and measured  $\langle \Delta r^2(t) \rangle$ for both  active and  passive spheres. We clearly find that the passive spheres are trapped within their cages as $\langle \Delta r^2(t) \rangle \ll \sigma^2$, whereas the single active hard sphere can  jump out of its cage, although it remains a  rare event. If we increase the fraction of active spheres to $5\%$, we find that the mobility of both the passive and active spheres increases significantly resulting in  long-time diffusion coefficients $D_{\mathrm{L}}/D_0 \simeq 0.02$ and 0.0625 for the passive and active particles, respectively. We thus conclude that  a motorized particle can occasionally break out of its cage by its own self-propelling force, but there is also a collective effect, because the mobility is even more enhanced upon increasing the fraction of active spheres.

To shed light on the enhanced mobility with increased activity, we investigate the structure of the active hard-sphere system as a function of self-propulsion $f$. We calculate the static structure factor  $S(q)$ at $\phi=0.62$ and present our results in  Fig. \ref{fig3}b. We clearly observe that the main peak of  $S(q)$ shifts to higher wave vector $q$ and becomes broader upon increasing the self-propulsion $f$, which indicates that the average separation between neighboring particles becomes smaller,   but the distribution of interparticle distances gets broader. We thus conclude that the structure of an equilibrium glass of passive hard spheres is different from a non-equilibrium active hard-sphere system at the same packing fraction. However, we also find that the difference in structure decreases upon increasing $\phi$, and we speculate that the structure becomes similar in the limit of RCP as the system becomes fully jammed. The   observed structural changes  can be explained by the fact that particles are pushed closer together due to  the self-propulsion, thereby creating more  ``free volume'' for  other particles, which can then break out of their cages and the glass state can melt  into a non-equilibrium fluid phase.  A similar melting of hard-sphere glasses was also observed in systems, where the addition of depletants instead of self-propulsions leads to a clustering of the particles and a destruction of the cages~\cite{poon2002,eckert2002}. A similar picture emerged from a detailed simulation study on a 2-dimensional system of active hard disks, where it was found that active particles transiently stick to the neighbours found in the direction of their self-propulsion until the direction of the motion is changed due to Brownian rotation~\cite{berthier2013prl}.

\subsection*{Effect of rotational diffusion coefficient}
\begin{figure}[ht!]
\includegraphics[width = \textwidth]{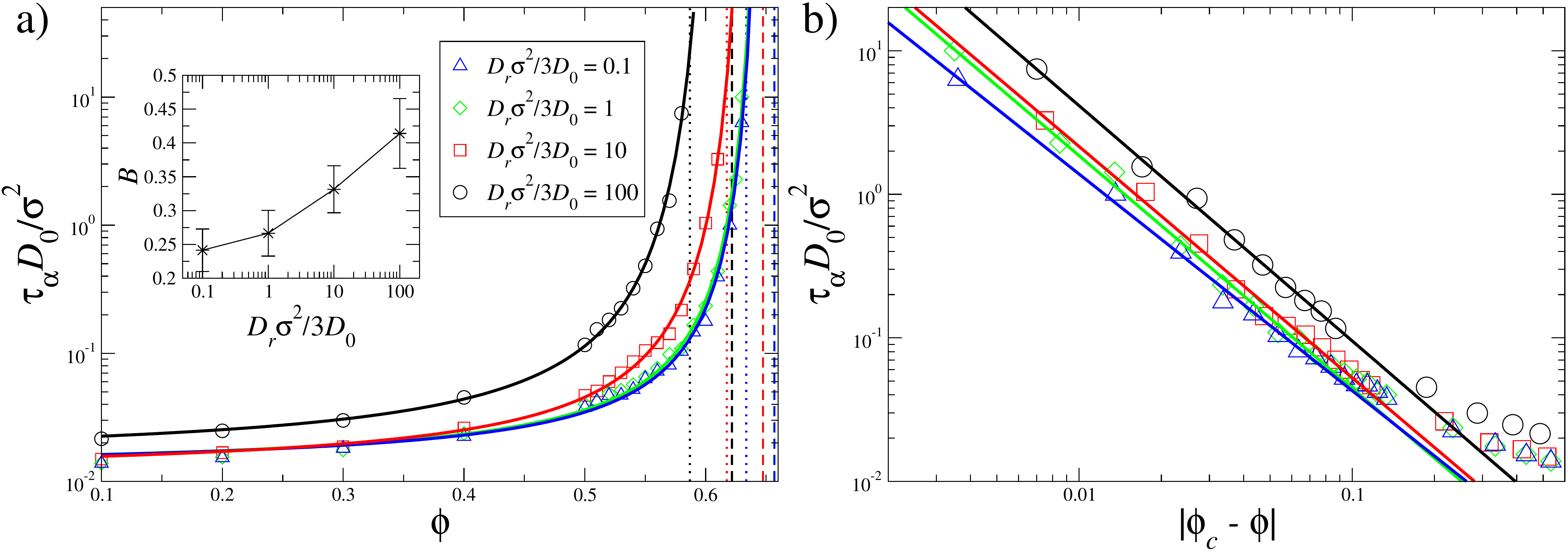}
\caption{\label{fig5} {\bf Effect of rotational diffusion on the glass transition.} (a) Structural relaxation time $\tau_\alpha$ as a function of packing fraction $\phi$ for active hard-sphere systems with a self-propulsion $f\sigma/k_\bb  T = 20$ and various rotational diffusion coefficients $D_{\mathrm{r}}$. The solid lines denote the VFT fits. The vertical dotted and dashed lines denote $\phi_\mathrm{c}$ and $\phi_0$ as obtained from the MCT and VFT fits, respectively. Note that  $\phi_\mathrm{c}$ and $\phi_0$ for $D_{\mathrm{r}}\sigma^2/3D_0 = 0.1$ and 1 are both very close to each other and hard to  distinguish in the figure.
Inset: the fragility index $B$ from the VFT fitting as a function of $D_{\mathrm{r}}$.
(b) shows the corresponding MCT fits (solid lines).
}
\end{figure}

In this work, we used a  rotational diffusion coefficient $D_{\mathrm{r}} = 3 D_{0}/\sigma^2$ according to the Stokes-Einstein relation.
To study the effect of particle rotations on the dynamics of the system, we perform EDBD simulations for a system of active hard spheres with $f\sigma/k_\bb  T = 20$ and rotational diffusion coefficients $D_{\mathrm{r}}$ varying from 0.3 to 300 $D_0/\sigma^2$. We  present the results in Fig.~\ref{fig5}. We observe clearly that the structural relaxation time increases monotonically with  packing fraction, and that the dynamics is well-described by  the VFT  and MCT predictions for all values of $D_{\mathrm{r}}$ that we studied. Note that $D_{\mathrm{r}} \sigma^2 / 3 D_{0}=1$ corresponds to the Stokes-Einstein relation.  In addition, we find that upon increasing the rotational diffusion coefficient $D_{\mathrm{r}}$, the relaxation dynamics slows down, the critical MCT and VFT packing fractions $\phi_\mathrm{c}$ and $\phi_0$ decrease, and the fragility index $B$ increases. More specifically, we find that the relaxation dynamics of active hard spheres approaches that of passive Brownian hard spheres in the limit of $D_{\mathrm{r}} \rightarrow \infty$ as the orientations of the self-propelling force at any time $t$ and $t'$ ($t' \neq t $) become uncorrelated thereby recovering the white noise in Brownian motion.

\section*{Discussion}
In conclusion, we performed extensive EDBD simulations of systems of self-propelled motorized hard spheres, and investigated the dynamical behavior of systems that are far out of equilibrium. We find that the relaxation dynamics can be sped up by orders of magnitude upon increasing the self-propulsion of the particles. This result provides an important step forward towards a better understanding of dense  packings as it allows us to investigate sphere packings with fluid-like dynamics at densities close to random close packing without crossing a glass transition. For instance, the structural relaxation time of  active hard spheres with a self-propulsion $f \sigma/k_\bb  T=80$ at a packing fraction as high as $\phi=0.63$  is similar to that of a passive hard-sphere system at $\phi=0.5$.

Moreover,  we show that  the  non-equilibrium MCT glass transition in a system of active hard spheres is pushed from  $~58 \%$ volume fraction in the case of passive hard spheres to packing fractions close to RCP upon increasing the self-propulsion. The non-equilibrium glass phase shows  characteristics like dynamical arrest, cooperative motion, and a diverging structural relaxation time that are reminiscent of the equilibrium glass transition. Thus, our findings strongly support recent theoretical predictions on the existence of a non-equilibrium glass transition in active systems~\cite{berthier2013}.  Consequently, as the packing fraction at which the system falls out of equilibrium depends on  self-propulsion, and the resulting glass phase can only be compressed further to its corresponding jammed state at infinite pressure,   we expect to find a range of jamming densities that depends on activity. In addition, our structure factor data  show that the structure of the equilibrium glass phase of passive hard spheres is different from that of active hard spheres, but we expect that the structure becomes similar in the limit of random close packing.

Additionally, we  show that the effect of a self-propulsion is more pronounced for  slower rotational diffusion, while the dynamics of active hard spheres becomes Brownian in the limit of an infinitely large rotational diffusion coefficient.

Our study  can also be  extended to  random packings of arbitrarily shaped particles, which will provide new insights in the concept of RCP as well as the glass transition of  these systems. It is also interesting to investigate if active particles can be used to unjam systems that are kinetically trapped in a glass phase in order to promote the formation of the  thermodynamically stable phase.  We  hope that our findings inspire future experimental investigations on  glassy dynamics of dense suspensions of active particles or of colloidal glasses that only contain a small amount of active particles. We  wish to note here  that most experimental studies so far have only been focussed on dilute instead of concentrated suspensions, and that most experimental systems correspond to much lower activities $f \sigma/k_\bb  T < 10$ \cite{howse2007,palacci2010,volpe2011,schwarz-linek2012,wilson2012} than studied here, but the light-activated colloids \cite{palacci2013}, the catalytic Janus particles \cite{baraban2012}, and the particles with a artificial magnetic flagella \cite{dreyfus2005} are capable of producing self-propulsions as high as $f \sigma/k_\bb  T \simeq 20$, 50, 80, respectively, which should have a profound effect on the relaxation dynamics in concentrated systems.

\section*{Methods} 
\subsection*{Simulation details}
We perform event-driven Brownian Dynamics simulations ~\cite{scala2007,scala2013} of a system of self-propelled hard spheres.
To avoid crystallization we study a suspension of $N=2000$ hard spheres with a Gaussian size polydispersity of
$8\%$ and a mean diameter of $\sigma$. Even though the particles are driven and energy is continuously supplied to the system, the temperature  $T$ is kept fixed  as  the solvent acts as a heat bath.
The overdamped motion of particle $i$ with position $\mathbf{r}_i$ and orientation $\mathbf{\hat{u}}_i$ is given by	
\begin{equation}
	\mathbf{\dot{r}}_i(t) = \frac{D_{0,i}}{k_\bb  T} \left \lbrack -\nabla_i U(t) + \mathbf{\xi}_i(t) + f \mathbf{\hat{u}}_i(t) \right \rbrack,
\end{equation}
where the potential energy $U=\sum_{i<j} U_{\mathrm{HS}}(r_{ij})$ is the sum of excluded-volume interactions between all hard spheres, and $D_{0,i} \propto 1/\sigma_i$ is the short-time self diffusion coefficient of particle $i$. A stochastic force  with zero mean, $\mathbf{\xi}_i(t)$, describes the  collisions with the solvent molecules, and satisfies $\langle \mathbf{\xi}_i(t)\mathbf{\xi}_j^T(t') \rangle = 2 (k_\bb  T)^2\mathbf{1}\delta_{ij}\delta(t-t')/D_{0,i}$ with $\mathbf{1}$ the identity matrix. In addition, the self-propulsion of particle $i$ is described by a  constant force $f$ in the direction $\mathbf{\hat{u}}_i(t)$ at time $t$.

%

\clearpage
\begin{table}[ht!]
\caption{The critical packing fraction $\phi_0$ and fit parameter $B$ as obtained from the VFT fit ($\tau_\alpha \sim \exp[B\phi_0 / (\phi_0 - \phi)]$) of the   structural relaxation time $\tau_\alpha$, and the critical packing fraction $\phi_\mathrm{c}$ and fit parameter $\gamma$ as extracted from the MCT fit ($\tau_\alpha \sim |\phi_\mathrm{c} - \phi|^{-\gamma}$) for active hard spheres with varying self-propulsions $f\sigma/k_\bb  T$. The errors are determined from a $95\%$ confidence interval for the fits.}\label{tab1}
\begin{tabular}{ccccc}
\hline
$f\sigma/k_\bb  T$ & \multicolumn{2}{c}{VFT fitting} & \multicolumn{2}{c}{MCT fitting}  \\
{}   & $B$   & $\phi_0$    & $\gamma$   & $\phi_\mathrm{c}$\\
\hline
0       &       $0.4888 \pm 0.0199$     &       $0.6201 \pm 0.0048$     &       $1.844 \pm 0.551$       & $0.5811 \pm 0.0069$\\
10      &       $0.3551 \pm 0.0480$ &   $0.6503 \pm 0.0054$ &   $1.631 \pm 0.277$       & $0.6175 \pm 0.0065$\\
20      &       $0.2666 \pm 0.0338$ &   $0.6568 \pm 0.0035$ &   $1.620 \pm 0.220$       & $0.6335 \pm 0.0017$\\
80      &       $0.1681 \pm 0.0176$ &   $0.6573 \pm 0.0019$ &   $1.546 \pm 0.246$       & $0.6423 \pm 0.0013$\\
800     &       $0.1487 \pm 0.0288$ &   $0.6573 \pm 0.0035$ &   $1.442 \pm 0.291$       & $0.6426 \pm 0.0034$\\
\hline
\end{tabular}
\end{table}

\clearpage
\clearpage
\clearpage
\clearpage
\clearpage

\end{document}